\documentclass[reprint,amsmath,amssymb,aps,pra,superscriptaddress]{revtex4-1}
\usepackage{amsmath,epsfig,wrapfig,graphicx,color,bm}

\begin{document}
\title{One- and three-dimensional quantum phase transitions and anisotropy in Rb$_{2}$Cu$_{2}$Mo$_{3}$O$_{12}$}

\author{S.~Hayashida}
\email{shoheih@phys.ethz.ch}
\affiliation{Laboratory for Solid State Physics, ETH Z{\"u}rich, 8093 Z{\"u}rich, Switzerland}
\author{D.~Blosser}
\affiliation{Laboratory for Solid State Physics, ETH Z{\"u}rich, 8093 Z{\"u}rich, Switzerland}
\author{K.~Yu.~Povarov}
\affiliation{Laboratory for Solid State Physics, ETH Z{\"u}rich, 8093 Z{\"u}rich, Switzerland}
\author{Z.~Yan}
\affiliation{Laboratory for Solid State Physics, ETH Z{\"u}rich, 8093 Z{\"u}rich, Switzerland}
\author{S.~Gvasaliya}
\affiliation{Laboratory for Solid State Physics, ETH Z{\"u}rich, 8093 Z{\"u}rich, Switzerland}
\author{A.~N.~Ponomaryov}
\affiliation{Dresden High Magnetic Field Laboratory (HLD-EMFL), Helmholtz-Zentrum Dresden-Rossendorf, 01328 Dresden, Germany}
\author{S.~A.~Zvyagin}
\affiliation{Dresden High Magnetic Field Laboratory (HLD-EMFL), Helmholtz-Zentrum Dresden-Rossendorf, 01328 Dresden, Germany}
\author{A.~Zheludev}
\email{zhelud@ethz.ch}\homepage{http://www.neutron.ethz.ch/}
\affiliation{Laboratory for Solid State Physics, ETH Z{\"u}rich, 8093 Z{\"u}rich, Switzerland}

\date{\today}

\begin{abstract}
Single crystal samples of the frustrated quasi one-dimensional quantum magnet Rb$_{2}$Cu$_{2}$Mo$_{3}$O$_{12}$ are investigated by magnetic, thermodynamic,  and electron spin resonance (ESR) measurements. Quantum phase transitions between the gapped, magnetically ordered and fully saturated phases are observed. Surprisingly, the former has a distinctive three-dimensional character, while the latter is dominated by one-dimensional quantum spin fluctuations. The entire $H$-$T$ phase diagram is mapped out and found to be substantially anisotropic. In particular, the lower critical fields differ by over 50\% depending on the direction of applied field, while the upper ones are almost isotropic, as is the magnetization above saturation. The ESR spectra are strongly dependent on field orientation and point to a helical structure with a rigidly defined spin rotation plane.
\end{abstract}

\maketitle
\section{Introduction}
Frustrated $S=1/2$ spin chains with competing nearest-neighbor $J_{1}$ and next-nearest-neighbor $J_{2}$ interactions are known to realize a panoply of exotic quantum phase such as chiral spin liquids~\cite{Chubukov1991, Kolezhuk2005,Hikihara2008,Sudan2009,Furukawa2010,Furukawa2012,Ueda2014_1,Ueda2014_2}, spin nematics~\cite{Chubukov1991, Kecke2007,Sudan2009,Zhitomirsky2010,Sato2013} and
spin density waves~\cite{Sudan2009,Sato2013,Nishimoto2015}.
One of the most intriguing species of current interest is the linear chain molybdate Rb$_{2}$Cu$_{2}$Mo$_{3}$O$_{12}$~\cite{Solodovnikov1997,Hase2004,Hase2005}.
It is believed to feature a competition of ferromagnetic $J_{1}=-138$~K and antiferromagnetic $J_{2}=51$~K $(J_{2}/|J_{1}|=0.37)$ interactions~\cite{Hase2004,Hase2005}. 
The ground state is a spin singlet with a gap $\Delta\sim 1.6$~K in the excitation spectrum~\cite{Yasui2014}. 
Powder samples have been extensively investigated by magnetic and dielectric  measurements~\cite{Yasui2013_1,Yasui2013_2,Reynolds2019}, high pressure studies~\cite{Kuroe2006,Hamasaki2007}, NMR~\cite{Yagi2017,Matsui2017}, neutron scattering~\cite{Tomiyasu2009,Reynolds2019} and muon spin relaxation measurements~\cite{Kawamura2018,Reynolds2019}.
The most interesting property is ferroelectric behavior that appears below 8~K despite the absence of any conventional magnetic order~\cite{Yasui2013_1,Yasui2013_2,Reynolds2019}. 
It was attributed to emergence of spin-chirality and the so-called  spin-current or inverse Dzyaloshinskii-Moriya mechanism~\cite{Katsura2005,Mostovoy2006,Jia2007,Xiang2007}.

Unfortunately, a lack of single crystal samples severely hampers any further experimental progress. 
For instance, almost nothing is known about the magnetic phase diagram~\cite{Matsui2017} or magnetic anisotropy in the system. 
In a breakthrough, we hereby present comprehensive magnetothermodynamic and electron spin resonance (ESR) measurements on Rb$_{2}$Cu$_{2}$Mo$_{3}$O$_{12}$ {\em single crystals}. 
We map out the entire $H$-$T$ phase diagram down to $0.1$~K in temperature and up to full saturation in magnetic field and find it to be highly anisotropic. 
The most intriguing result is that while the field-induced ordering transition is of a distinct three-dimensional character, the quantum phase transition at saturation is entirely dominated by one-dimensional fluctuations.

\section{Experimental details}
\begin{figure}[tbp]
\includegraphics[scale=1]{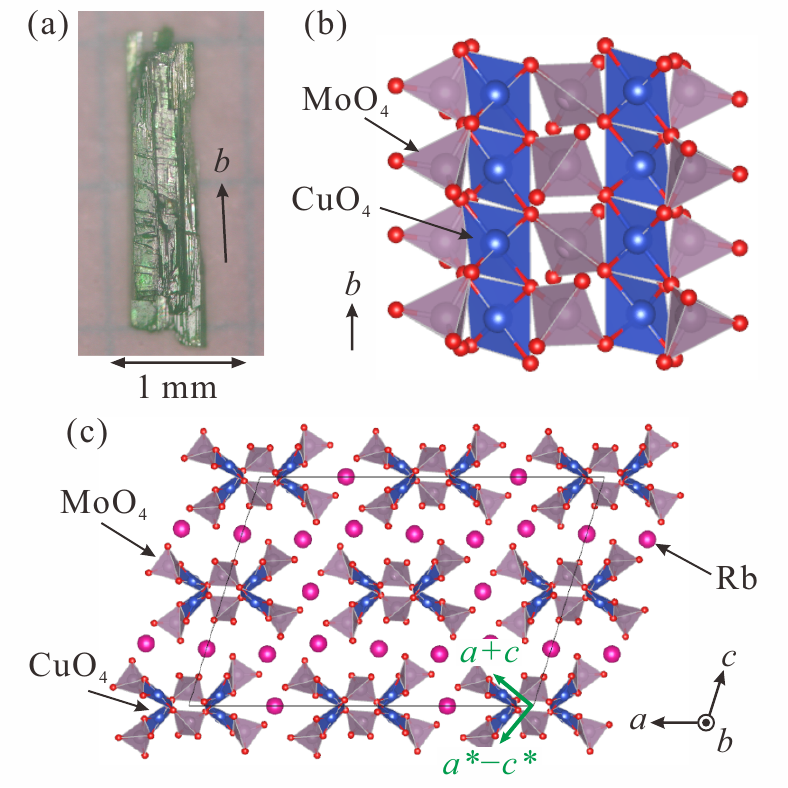}
\caption{(a) Typical single crystal Rb$_{2}$Cu$_{2}$Mo$_{3}$O$_{12}$ sample used in this work. (b),(c) Schematic view of the crystal structures of Rb$_{2}$Cu$_{2}$Mo$_{3}$O$_{12}$ (monoclinic, space-group $C2/c$).
}
\label{fig1}
\end{figure}

Single crystal samples of Rb$_{2}$Cu$_{2}$Mo$_{3}$O$_{12}$ with typical mass $0.1$~mg were grown by a spontaneous crystallization in a flux method~\cite{Solodovnikov1997}.
Green transparent crystals were obtained as shown in Fig.~\ref{fig1}(a).
The crystal structure [monoclinic $C2/c$, $a=27.698(7)$~{\AA}, $b=5.1010(15)$~{\AA}, $c=19.291(3)$~{\AA}, $\alpha=90^{\circ}$, $\beta=107.31(3)^{\circ}$, $\gamma=90^{\circ}$] was validated
using single-crystal x-ray diffraction on a Bruker APEX-II instrument. 
It was found to be totally consistent with the previous report~\cite{Solodovnikov1997}. 
The magnetic properties are due to $S=1/2$ Cu$^{2+}$ cations. 
As illustrated in Figs.~\ref{fig1}(b) and \ref{fig1}(c), CuO$_{4}$ plaquettes form one-dimensional (1D) chains along the crystallographic $b$ axis.
The individual chains are paired via MoO$_{4}$ bridges. The resulting chain-pairs are separated from one another by the Rb$^{+}$ ions.

Bulk measurements were carried out using the $^{3}$He-$^{4}$He dilution refrigerator insert for the Quantum Design physical property measurement system (PPMS).
Heat capacity data were collected on a standard Quantum Design relaxation calorimetry option.
The magnetic field was applied either along the crystallographic $a^{*}-c^{*}$ (transverse) or $b$ (longitudinal) directions, correspondingly.
Heat capacity was measured in the range of $0.1$~K~$\leq T \leq1.7$~K and $0$~T~$\leq H \leq 14$~T without background subtraction.
Magnetization measurements were carried out on a home-made Faraday force magnetometer.
Field scans were collected at $0.1$~K and $3$~K up to $14$~T, with the magnetic field along the $a^{*}-c^{*}$, $a+c$ and $b$ axes, respectively.
The absolute value of magnetization was obtained in a calibration measurement at $3$~K using the vibrating sample magnetometer (VSM)
for the PPMS. 
ESR measurements were performed using a $16$~T transmission-type ESR spectrometer, similar to that described in Ref.~\cite{Zvyagin2004}.
We measured field scans at $1.4$~K at several frequencies.
The experiments were set in the Faraday or Voigt configurations with the magnetic field applied along the
$a^{*}-c^{*}$ and $b$ axes, respectively.

\section{Results and Discussion}
\subsection{Heat capacity}
\begin{figure}[tbp]
\includegraphics[scale=1]{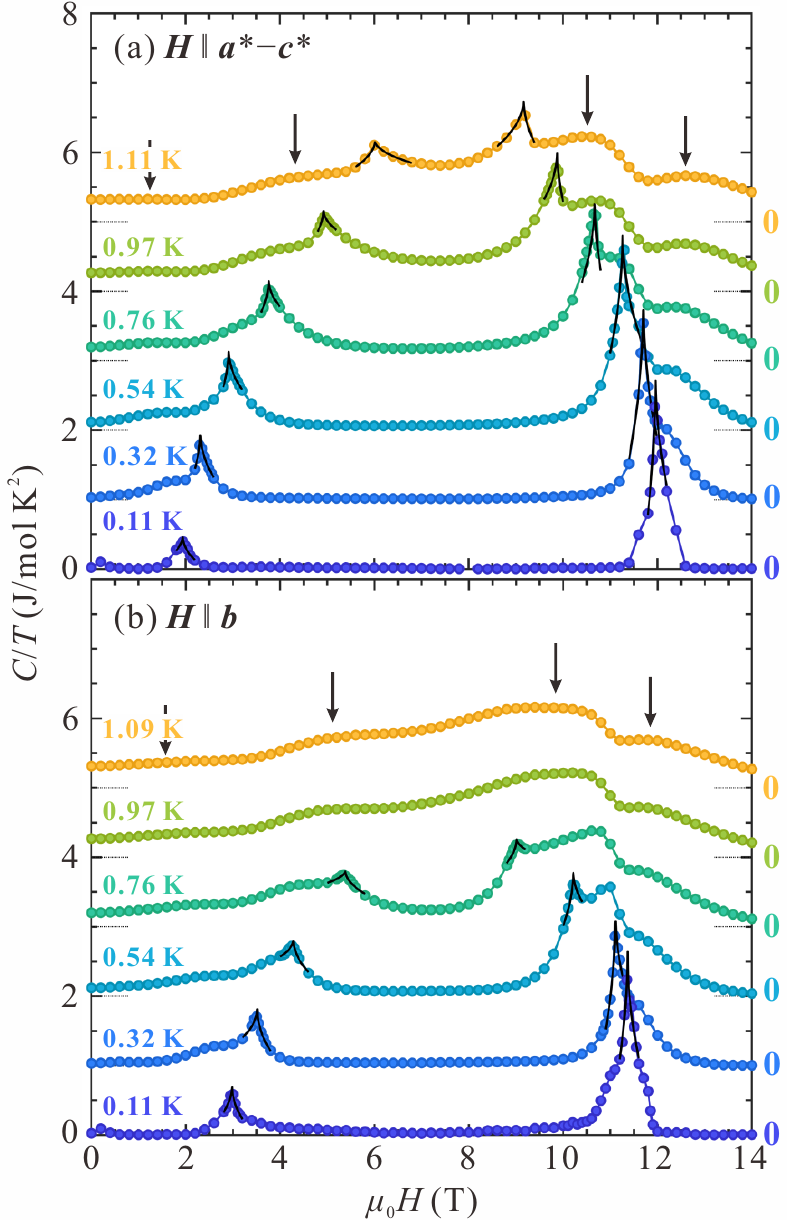}
\caption{Typical measured constant-temperature field scans of heat capacity in Rb$_{2}$Cu$_{2}$Mo$_{3}$O$_{12}$ for the transverse (a) and longitudinal (b) field geometries. Additional solid curves in the vicinity of sharp peaks are empirical power-law fits used to pinpoint the transition fields. Arrows indicate positions of additional broad features in the data, as described in the text.
For visibility the scans are offset by $0.1$~J/mol K$^2$ relative to one another.}
\label{fig2}
\end{figure}
Typical measured field dependencies of heat capacity are shown in Fig.~\ref{fig2}.
For the transverse and longitudinal field geometries, pairs of sharp lambda-anomalies are observed below 1.2~K and 0.9~K, correspondingly. 
We attribute these to a phase transition from the paramagnetic state to three-dimensional (3D) long-range
order (LRO). In the low temperature limit the lower anomaly corresponds to a closure of the spin gap and the upper one to saturation. At higher
temperatures the anomalies come closer together tracing a typical ''dome'' shape on the phase diagram \cite{Zapf2014}. 
For the two orientations studied, the critical fields at $0.1$~K are $H_{c1,\bot}=1.9$~T, $H_{c2,\bot}=12.0$~T, $H_{c1,\|}=3.0$~T, and $H_{c2,\|}=11.4$~T for the transverse and longitudinal orientations, respectively. 
Strikingly, the upper critical fields are almost the same in the two geometries, but the lower ones differ by over 50\%.

In addition to the sharp peaks, the field scans of specific heat show broad but prominent double-hump features (arrows in Fig.~\ref{fig2}), particularly near saturation. Such features are typical of $d=1$, $z=2$ quantum criticality~\cite{Korepin1990,Sachdev1994} and are also observed in other 1D materials magnets~\cite{Ruegg2008,Breunig2017,Blosser2018}.

\begin{figure}[tbp]
\includegraphics[scale=1]{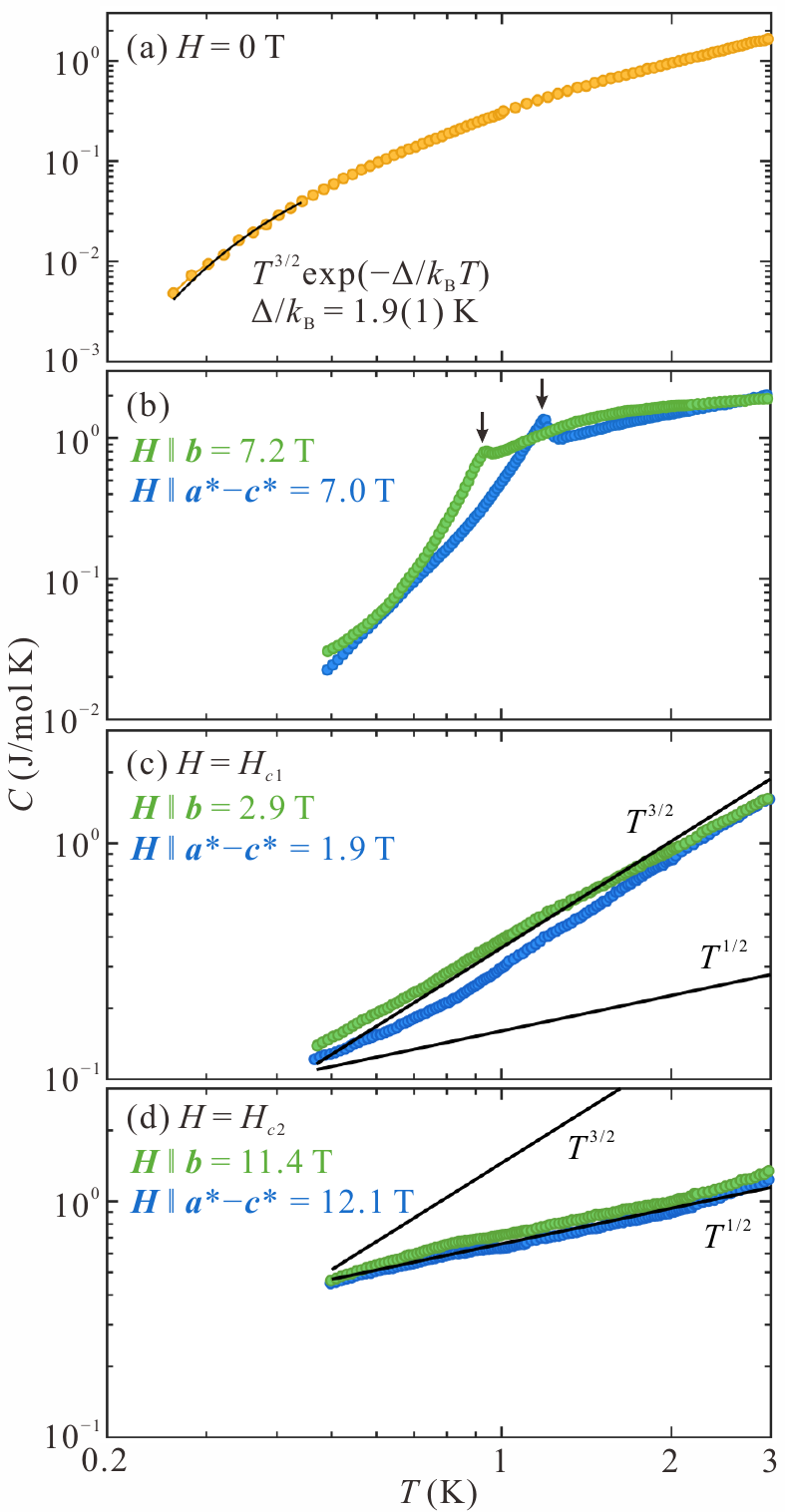}
\caption{Symbols: temperature scans of the heat capacity measured in  Rb$_{2}$Cu$_{2}$Mo$_{3}$O$_{12}$ at different fields. (a)  Zero applied field.
The solid line is a fit of an exponentially activated form, as described in the text.  (b) Intermediate fields. Arrows indicate lambda anomalies that represent long range magnetic ordering. (c),(d) Magnetic fields $H_{c1}$ and $H_{c2}$ corresponding to quantum phase transitions at gap closure and full saturation, respectively. The solid lines are power laws with exponents of $3/2$ and $1/2$.}
\label{fig3}
\end{figure}
\begin{figure}[tbp]
\includegraphics[scale=1]{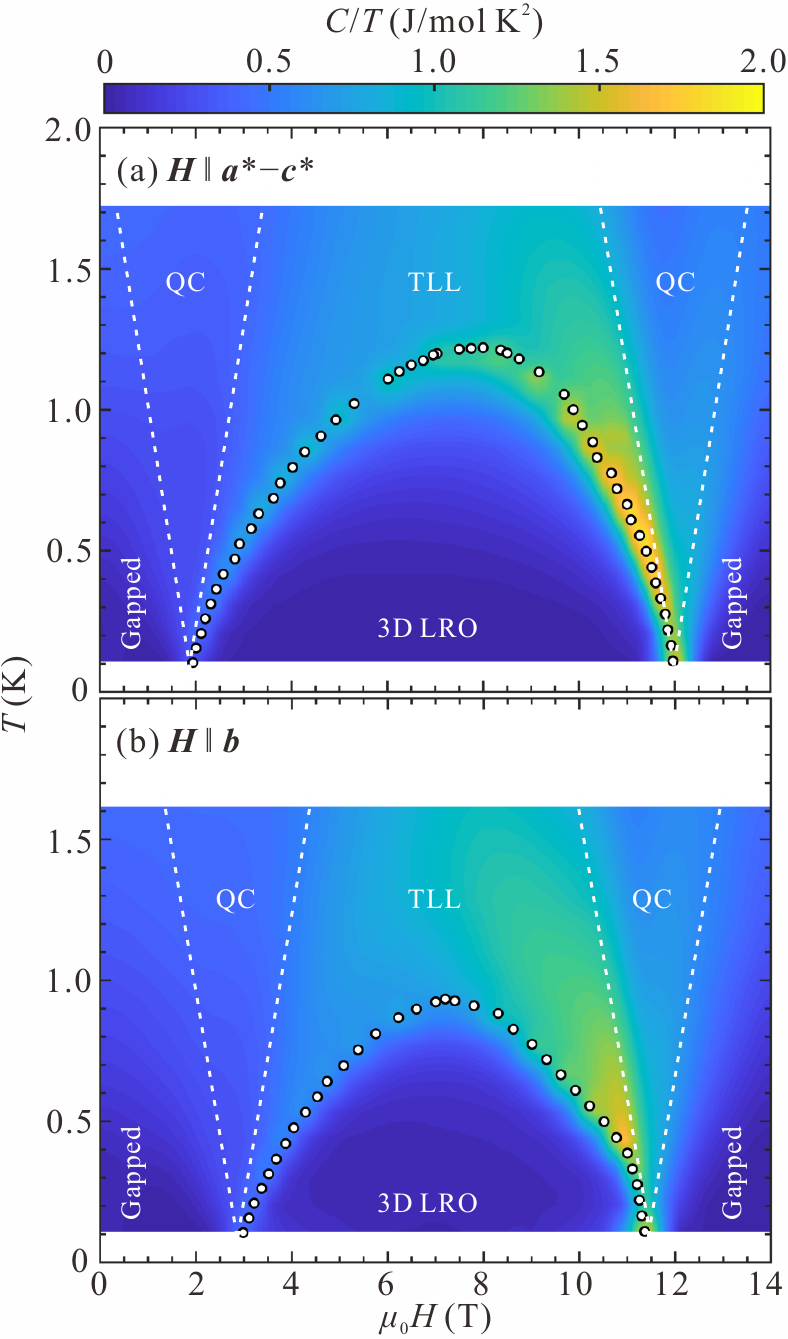}
\caption{Symbols: magnetic phase diagrams of Rb$_{2}$Cu$_{2}$Mo$_{3}$O$_{12}$ in the transverse (a) and longitudinal (b) geometries.
The backgrounds show corresponding false color maps of $C(T,H)/T$. The phase regions are labeled as follows: three-dimensional long-range order (3D LRO),  gapped, Tomonaga-Luttinger spin liquid (TLL), and quantum critical regime (QC).
Dashed lines are guides for the eye.
}
\label{fig4}
\end{figure}

The temperature dependence of the heat capacity at zero field is shown in Fig.~\ref{fig3}(a).  
Note the logarithmic  scales on both axes.
In full agreement with a gapped excitation spectrum, the low temperature behavior is exponentially activated.
Assuming that a gapped quadratic one-dimensional dispersion relation for the low-lying excitations, 
the specific heat is given by $C_V\propto T^{3/2}\exp(-\Delta/k_\mathrm{B}T)$ in the low-temperature limit $k_\mathrm{B}T\ll \Delta$~\cite{Troyer1994,Hong2010}.
The activation temperature $\Delta/k_\mathrm{B}$ is determined to be $1.9(1)$~K, which is roughly consistent with the lower critical field values.

In applied fields, across the domes of the 3D ordered phase, constant-$H$ temperature scans of specific heat show distinct lambda anomalies at the phase boundary [Fig.~\ref{fig3}(b)]. 
Most telling is temperature scans at precisely $H_{c1}$ and $H_{c2}$ shown in Figs.~\ref{fig3}(c) and \ref{fig3}(d) respectively. 
They reveal that, regardless of field geometry, the nature of the corresponding quantum critical points is markedly different. 
In general, for field-induced quantum phase transitions in gapped spin systems where magnons have a quadratic dispersion, we expect $C_V\propto T^{d/2}$~\cite{ Continentino2017}. 
In this context, the lower transition behaves much as we would expect for 3D ordering, with $C_V\propto T^{3/2}$. 
In contrast, the criticality at $H_{c2}$ must be dominated by one-dimensional fluctuations, as clearly $C_V\propto T^{1/2}$ provides a much better description of the data.

The bulk of the measured specific heat data was used to reconstruct the magnetic phase diagrams shown in Fig.~\ref{fig4} over false-color heat capacity plots. 
The symbols are boundaries of the 3D-ordered phase traced by the lambda anomalies in constant-$T$ (Fig.~\ref{fig2}) or constant-$H$ [Fig.~\ref{fig3}(b)] scans. 
The measured phase diagram is generally consistent with the one measured in powder samples with NMR~\cite{Matsui2017}.
The domes of the ordered phase are markedly anisotropic, with the maximal ordering temperature visibly suppressed in the longitudinal field configuration. 
For that geometry there is also a peculiar kink on the phase boundary at $H\sim 11$~T and $T\sim0.5$~K.
Also noteworthy is that the specific heats are strongly enhanced just above the upper boundary  for both geometries.

\subsection{Magnetization}
\begin{figure}[tbp]
\includegraphics[scale=1]{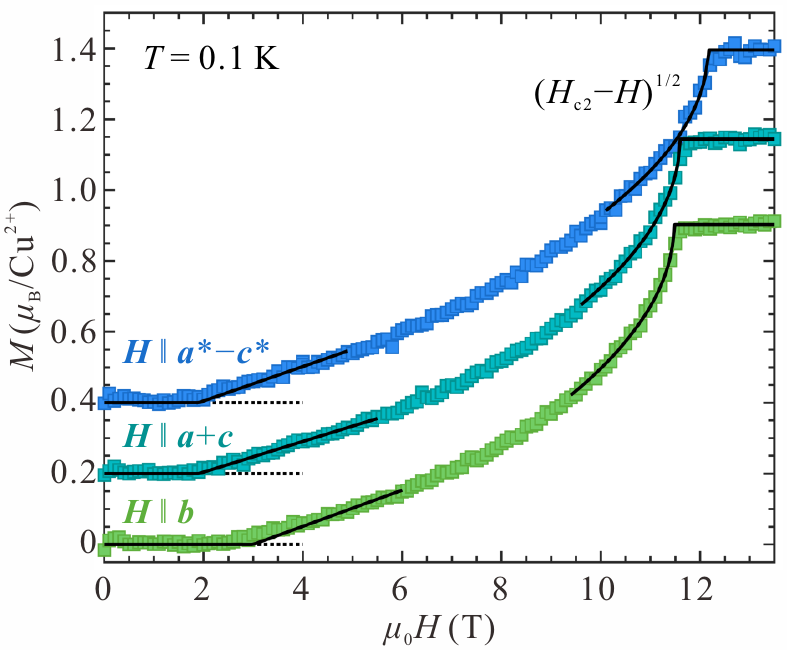}
\caption{Magnetization curves measured in Rb$_{2}$Cu$_{2}$Mo$_{3}$O$_{12}$ at $T=0.1$~K in fields applied along three crystallographic directions (symbols). Solid lines are linear and square-root fits to the data in the vicinity of $H_{c1}$ and $H_{c2}$, respectively. Two data sets are offset by $0.2$~$\mu_\mathrm{B}$ and  $0.4$~$\mu_\mathrm{B}$ for visibility.
}
\label{fig5}
\end{figure}

Magnetization curves measured in Rb$_{2}$Cu$_{2}$Mo$_{3}$O$_{12}$ at $T=0.1$~K for three field geometries  are shown in Fig.~\ref{fig5}.
The nonmagnetic ground state below $2$~T is confirmed.
The transition fields at about $2$~T and at about $11.5$~T coincide with the critical fields $H_{c1}$ and $H_{c2}$ as measured with heat capacity.
The saturated magnetizations are almost isotropic for the three orientations, and they are
$0.99(1)$~$\mu_{\rm B}$ for ${\bm H}\|({\bm a}^{*}-{\bm c}^{*})$, $0.95(1)$~$\mu_{\rm B}$ for ${\bm H}\|({\bm a}+{\bm c})$, and $0.90(1)$~$\mu_{\rm B}$ for ${\bm H}\|{\bm b}$. 
In all geometries the most striking feature is the different behaviors of magnetization in the vicinity of the critical fields. 
Near $H_{c2}$ we see a distinct square-root approach to saturation. 
This is a hallmark of a $d=1$ $z=2$ quantum phase transition and is typical for one-dimensional Heisenberg spin systems~\cite{Bonner1964}. 
In contrast, near $H_{c1}$, to within experimental noise, there is no sign of a square root singularity. Instead, we see a linear increase of magnetization. 
This mean-field behavior is typical of $d=3$ $z=2$ quantum criticality often referred to as a BEC of magnons \cite{Batyev1984,Giamarchi1999}.

\subsection{ESR}
\begin{figure}[tbp]
\includegraphics[scale=1]{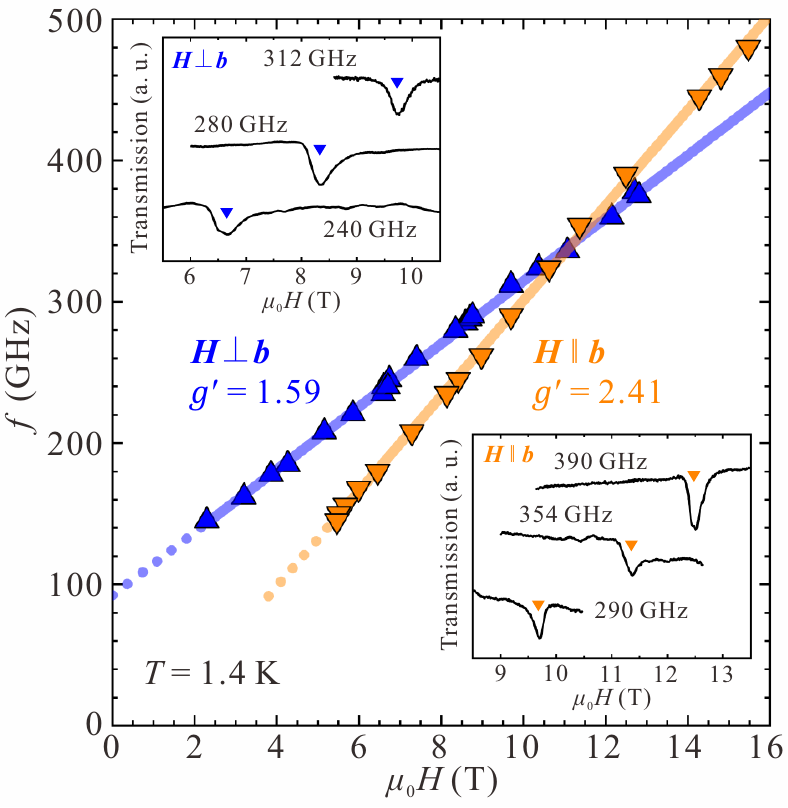}
\caption{Frequency-field diagram ESR excitations measured at $T=1.4$~K. Solid lines are the fits
of the linear function $\Delta+g^{\prime}\mu_{\rm B}H/h$. 
The insets show typical raw ESR spectra with characteristic transmission dips marking the resonance frequency.
}
\label{fig6}
\end{figure}

 In low-temperature ESR experiments, in fields applied in either the longitudinal or transverse geometry, we observed a single resonance mode in all cases, as shown in the insets of Fig.~\ref{fig6}. The main panel shows the  measured field dependence of the resonance frequencies.
In both geometries the behavior is linear in the accessible measurement range.
Linear fits with  $f=\Delta+g^{\prime}\mu_{\rm B}H/h$ yield $\Delta=92.2$~GHz, $g'=1.59$ for
${\bm H}\bot {\bm b}$
 and $\Delta=-36.5$~GHz, $g'=2.41$ for ${\bm H}\|{\bm b}$.
The effective $g^{\prime}$-factor is very different for the two geometries. 
They clearly do {\em not} correspond to the $g$ factor of Rb$_{2}$Cu$_{2}$Mo$_{3}$O$_{12}$ which, based on the values of saturation magnetization discussed above, is very isotropic.
One possible origin of this effect is that the ordered state is a spin spiral (helimagnet).
In such systems the effective $g^{\prime}$ factor with applied field parallel to the spin rotation plane is known to be larger than in a transverse orientation.
The ESR frequency dependence that we measure in Rb$_{2}$Cu$_{2}$Mo$_{3}$O$_{12}$ is indeed qualitatively similar to that seen in quasi-1D helimagnets such as LiCu$_{2}$O$_{2}$~\cite{Svistov2010}, Li$_{2}$ZrCuO$_{4}$~\cite{Fujita2014}, and LiCuVO$_{4}$~\cite{Prozorova2016}. 
The key point, unlike in those materials, is that our ESR measurements at $1.4$~K are above the ordering transition. 
This suggests that at $1.4$~K one-dimensional spiral correlations are already well established in Rb$_{2}$Cu$_{2}$Mo$_{3}$O$_{12}$, or may even be a signature of three-dimensional chiral order in the system
\cite{Kolezhuk2005,Hikihara2008,Sudan2009}.


\section{Conclusion}
Magnetic, thermodynamic and ESR measurements enabled by single crystal samples provide several important clues to the nature of quantum magnetism in Rb$_{2}$Cu$_{2}$Mo$_{3}$O$_{12}$: (i) There is a huge anisotropy of lower critical fields, while the upper critical fields and saturation magnetizations are almost isotropic. 
(ii) At the lower critical field the quantum phase transition is a three-dimensional one with typical mean field behavior. 
The saturation transition, on the other hand, is dominated by one dimensional quantum fluctuations.
(iii) Even at temperatures above the top of the LRO dome there is an indirect sign of strong helimagnetic spin correlations along the $b$ axis. 
Magnetic anisotropy is thus a key ingredient in the rich physics of Rb$_{2}$Cu$_{2}$Mo$_{3}$O$_{12}$.

\section*{Acknowledgements}
This work was supported by Swiss National Science Foundation under Division II, and
Deutsche Forschungsgemeinschaft  (DFG) through ZV 6/2-2 and SFB 1143.
We acknowledge the support of the HLD at HZDR, member of the European Magnetic Field Laboratory (EMFL).

\begin{thebibliography}{45}%
\makeatletter
\providecommand \@ifxundefined [1]{%
 \@ifx{#1\undefined}
}%
\providecommand \@ifnum [1]{%
 \ifnum #1\expandafter \@firstoftwo
 \else \expandafter \@secondoftwo
 \fi
}%
\providecommand \@ifx [1]{%
 \ifx #1\expandafter \@firstoftwo
 \else \expandafter \@secondoftwo
 \fi
}%
\providecommand \natexlab [1]{#1}%
\providecommand \enquote  [1]{``#1''}%
\providecommand \bibnamefont  [1]{#1}%
\providecommand \bibfnamefont [1]{#1}%
\providecommand \citenamefont [1]{#1}%
\providecommand \href@noop [0]{\@secondoftwo}%
\providecommand \href [0]{\begingroup \@sanitize@url \@href}%
\providecommand \@href[1]{\@@startlink{#1}\@@href}%
\providecommand \@@href[1]{\endgroup#1\@@endlink}%
\providecommand \@sanitize@url [0]{\catcode `\\12\catcode `\$12\catcode
  `\&12\catcode `\#12\catcode `\^12\catcode `\_12\catcode `\%12\relax}%
\providecommand \@@startlink[1]{}%
\providecommand \@@endlink[0]{}%
\providecommand \url  [0]{\begingroup\@sanitize@url \@url }%
\providecommand \@url [1]{\endgroup\@href {#1}{\urlprefix }}%
\providecommand \urlprefix  [0]{URL }%
\providecommand \Eprint [0]{\href }%
\providecommand \doibase [0]{http://dx.doi.org/}%
\providecommand \selectlanguage [0]{\@gobble}%
\providecommand \bibinfo  [0]{\@secondoftwo}%
\providecommand \bibfield  [0]{\@secondoftwo}%
\providecommand \translation [1]{[#1]}%
\providecommand \BibitemOpen [0]{}%
\providecommand \bibitemStop [0]{}%
\providecommand \bibitemNoStop [0]{.\EOS\space}%
\providecommand \EOS [0]{\spacefactor3000\relax}%
\providecommand \BibitemShut  [1]{\csname bibitem#1\endcsname}%
\let\auto@bib@innerbib\@empty
\bibitem [{\citenamefont {Chubukov}(1991)}]{Chubukov1991}%
  \BibitemOpen
  \bibfield  {author} {\bibinfo {author} {\bibfnamefont {A.~V.}\ \bibnamefont
  {Chubukov}},\ }\href {\doibase 10.1103/PhysRevB.44.4693} {\bibfield
  {journal} {\bibinfo  {journal} {Phys. Rev. B}\ }\textbf {\bibinfo {volume}
  {44}},\ \bibinfo {pages} {4693} (\bibinfo {year} {1991})}\BibitemShut
  {NoStop}%
\bibitem [{\citenamefont {Kolezhuk}\ and\ \citenamefont
  {Vekua}(2005)}]{Kolezhuk2005}%
  \BibitemOpen
  \bibfield  {author} {\bibinfo {author} {\bibfnamefont {A.}~\bibnamefont
  {Kolezhuk}}\ and\ \bibinfo {author} {\bibfnamefont {T.}~\bibnamefont
  {Vekua}},\ }\href {\doibase 10.1103/PhysRevB.72.094424} {\bibfield  {journal}
  {\bibinfo  {journal} {Phys. Rev. B}\ }\textbf {\bibinfo {volume} {72}},\
  \bibinfo {pages} {094424} (\bibinfo {year} {2005})}\BibitemShut {NoStop}%
\bibitem [{\citenamefont {Hikihara}\ \emph {et~al.}(2008)\citenamefont
  {Hikihara}, \citenamefont {Kecke}, \citenamefont {Momoi},\ and\ \citenamefont
  {Furusaki}}]{Hikihara2008}%
  \BibitemOpen
  \bibfield  {author} {\bibinfo {author} {\bibfnamefont {T.}~\bibnamefont
  {Hikihara}}, \bibinfo {author} {\bibfnamefont {L.}~\bibnamefont {Kecke}},
  \bibinfo {author} {\bibfnamefont {T.}~\bibnamefont {Momoi}}, \ and\ \bibinfo
  {author} {\bibfnamefont {A.}~\bibnamefont {Furusaki}},\ }\href {\doibase
  10.1103/PhysRevB.78.144404} {\bibfield  {journal} {\bibinfo  {journal} {Phys.
  Rev. B}\ }\textbf {\bibinfo {volume} {78}},\ \bibinfo {pages} {144404}
  (\bibinfo {year} {2008})}\BibitemShut {NoStop}%
\bibitem [{\citenamefont {Sudan}\ \emph {et~al.}(2009)\citenamefont {Sudan},
  \citenamefont {L\"uscher},\ and\ \citenamefont {L\"auchli}}]{Sudan2009}%
  \BibitemOpen
  \bibfield  {author} {\bibinfo {author} {\bibfnamefont {J.}~\bibnamefont
  {Sudan}}, \bibinfo {author} {\bibfnamefont {A.}~\bibnamefont {L\"uscher}}, \
  and\ \bibinfo {author} {\bibfnamefont {A.~M.}\ \bibnamefont {L\"auchli}},\
  }\href {\doibase 10.1103/PhysRevB.80.140402} {\bibfield  {journal} {\bibinfo
  {journal} {Phys. Rev. B}\ }\textbf {\bibinfo {volume} {80}},\ \bibinfo
  {pages} {140402(R)} (\bibinfo {year} {2009})}\BibitemShut {NoStop}%
\bibitem [{\citenamefont {Furukawa}\ \emph {et~al.}(2010)\citenamefont
  {Furukawa}, \citenamefont {Sato},\ and\ \citenamefont
  {Onoda}}]{Furukawa2010}%
  \BibitemOpen
  \bibfield  {author} {\bibinfo {author} {\bibfnamefont {S.}~\bibnamefont
  {Furukawa}}, \bibinfo {author} {\bibfnamefont {M.}~\bibnamefont {Sato}}, \
  and\ \bibinfo {author} {\bibfnamefont {S.}~\bibnamefont {Onoda}},\ }\href
  {\doibase 10.1103/PhysRevLett.105.257205} {\bibfield  {journal} {\bibinfo
  {journal} {Phys. Rev. Lett.}\ }\textbf {\bibinfo {volume} {105}},\ \bibinfo
  {pages} {257205} (\bibinfo {year} {2010})}\BibitemShut {NoStop}%
\bibitem [{\citenamefont {Furukawa}\ \emph {et~al.}(2012)\citenamefont
  {Furukawa}, \citenamefont {Sato}, \citenamefont {Onoda},\ and\ \citenamefont
  {Furusaki}}]{Furukawa2012}%
  \BibitemOpen
  \bibfield  {author} {\bibinfo {author} {\bibfnamefont {S.}~\bibnamefont
  {Furukawa}}, \bibinfo {author} {\bibfnamefont {M.}~\bibnamefont {Sato}},
  \bibinfo {author} {\bibfnamefont {S.}~\bibnamefont {Onoda}}, \ and\ \bibinfo
  {author} {\bibfnamefont {A.}~\bibnamefont {Furusaki}},\ }\href {\doibase
  10.1103/PhysRevB.86.094417} {\bibfield  {journal} {\bibinfo  {journal} {Phys.
  Rev. B}\ }\textbf {\bibinfo {volume} {86}},\ \bibinfo {pages} {094417}
  (\bibinfo {year} {2012})}\BibitemShut {NoStop}%
\bibitem [{\citenamefont {Ueda}\ and\ \citenamefont
  {Onoda}(2014{\natexlab{a}})}]{Ueda2014_1}%
  \BibitemOpen
  \bibfield  {author} {\bibinfo {author} {\bibfnamefont {H.}~\bibnamefont
  {Ueda}}\ and\ \bibinfo {author} {\bibfnamefont {S.}~\bibnamefont {Onoda}},\
  }\href {\doibase 10.1103/PhysRevB.89.024407} {\bibfield  {journal} {\bibinfo
  {journal} {Phys. Rev. B}\ }\textbf {\bibinfo {volume} {89}},\ \bibinfo
  {pages} {024407} (\bibinfo {year} {2014}{\natexlab{a}})}\BibitemShut
  {NoStop}%
\bibitem [{\citenamefont {Ueda}\ and\ \citenamefont
  {Onoda}(2014{\natexlab{b}})}]{Ueda2014_2}%
  \BibitemOpen
  \bibfield  {author} {\bibinfo {author} {\bibfnamefont {H.}~\bibnamefont
  {Ueda}}\ and\ \bibinfo {author} {\bibfnamefont {S.}~\bibnamefont {Onoda}},\
  }\href {\doibase 10.1103/PhysRevB.90.214425} {\bibfield  {journal} {\bibinfo
  {journal} {Phys. Rev. B}\ }\textbf {\bibinfo {volume} {90}},\ \bibinfo
  {pages} {214425} (\bibinfo {year} {2014}{\natexlab{b}})}\BibitemShut
  {NoStop}%
\bibitem [{\citenamefont {Kecke}\ \emph {et~al.}(2007)\citenamefont {Kecke},
  \citenamefont {Momoi},\ and\ \citenamefont {Furusaki}}]{Kecke2007}%
  \BibitemOpen
  \bibfield  {author} {\bibinfo {author} {\bibfnamefont {L.}~\bibnamefont
  {Kecke}}, \bibinfo {author} {\bibfnamefont {T.}~\bibnamefont {Momoi}}, \ and\
  \bibinfo {author} {\bibfnamefont {A.}~\bibnamefont {Furusaki}},\ }\href
  {\doibase 10.1103/PhysRevB.76.060407} {\bibfield  {journal} {\bibinfo
  {journal} {Phys. Rev. B}\ }\textbf {\bibinfo {volume} {76}},\ \bibinfo
  {pages} {060407(R)} (\bibinfo {year} {2007})}\BibitemShut {NoStop}%
\bibitem [{\citenamefont {Zhitomirsky}\ and\ \citenamefont
  {Tsunetsugu}(2010)}]{Zhitomirsky2010}%
  \BibitemOpen
  \bibfield  {author} {\bibinfo {author} {\bibfnamefont {M.~E.}\ \bibnamefont
  {Zhitomirsky}}\ and\ \bibinfo {author} {\bibfnamefont {H.}~\bibnamefont
  {Tsunetsugu}},\ }\href {\doibase 10.1209/0295-5075/92/37001} {\bibfield
  {journal} {\bibinfo  {journal} {{EPL} (Europhysics Letters)}\ }\textbf
  {\bibinfo {volume} {92}},\ \bibinfo {pages} {37001} (\bibinfo {year}
  {2010})}\BibitemShut {NoStop}%
\bibitem [{\citenamefont {Sato}\ \emph {et~al.}(2013)\citenamefont {Sato},
  \citenamefont {Hikihara},\ and\ \citenamefont {Momoi}}]{Sato2013}%
  \BibitemOpen
  \bibfield  {author} {\bibinfo {author} {\bibfnamefont {M.}~\bibnamefont
  {Sato}}, \bibinfo {author} {\bibfnamefont {T.}~\bibnamefont {Hikihara}}, \
  and\ \bibinfo {author} {\bibfnamefont {T.}~\bibnamefont {Momoi}},\ }\href
  {\doibase 10.1103/PhysRevLett.110.077206} {\bibfield  {journal} {\bibinfo
  {journal} {Phys. Rev. Lett.}\ }\textbf {\bibinfo {volume} {110}},\ \bibinfo
  {pages} {077206} (\bibinfo {year} {2013})}\BibitemShut {NoStop}%
\bibitem [{\citenamefont {Nishimoto}\ \emph {et~al.}(2015)\citenamefont
  {Nishimoto}, \citenamefont {Drechsler}, \citenamefont {Kuzian}, \citenamefont
  {Richter},\ and\ \citenamefont {van~den Brink}}]{Nishimoto2015}%
  \BibitemOpen
  \bibfield  {author} {\bibinfo {author} {\bibfnamefont {S.}~\bibnamefont
  {Nishimoto}}, \bibinfo {author} {\bibfnamefont {S.-L.}\ \bibnamefont
  {Drechsler}}, \bibinfo {author} {\bibfnamefont {R.}~\bibnamefont {Kuzian}},
  \bibinfo {author} {\bibfnamefont {J.}~\bibnamefont {Richter}}, \ and\
  \bibinfo {author} {\bibfnamefont {J.}~\bibnamefont {van~den Brink}},\ }\href
  {\doibase 10.1103/PhysRevB.92.214415} {\bibfield  {journal} {\bibinfo
  {journal} {Phys. Rev. B}\ }\textbf {\bibinfo {volume} {92}},\ \bibinfo
  {pages} {214415} (\bibinfo {year} {2015})}\BibitemShut {NoStop}%
\bibitem [{\citenamefont {Solodovnikov}\ and\ \citenamefont
  {Solodovnikova}(1997)}]{Solodovnikov1997}%
  \BibitemOpen
  \bibfield  {author} {\bibinfo {author} {\bibfnamefont {S.~F.}\ \bibnamefont
  {Solodovnikov}}\ and\ \bibinfo {author} {\bibfnamefont {Z.~A.}\ \bibnamefont
  {Solodovnikova}},\ }\href {\doibase 10.1007/BF02763890} {\bibfield  {journal}
  {\bibinfo  {journal} {J. Struct. Chem.}\ }\textbf {\bibinfo {volume} {38}},\
  \bibinfo {pages} {765} (\bibinfo {year} {1997})}\BibitemShut {NoStop}%
\bibitem [{\citenamefont {Hase}\ \emph {et~al.}(2004)\citenamefont {Hase},
  \citenamefont {Kuroe}, \citenamefont {Ozawa}, \citenamefont {Suzuki},
  \citenamefont {Kitazawa}, \citenamefont {Kido},\ and\ \citenamefont
  {Sekine}}]{Hase2004}%
  \BibitemOpen
  \bibfield  {author} {\bibinfo {author} {\bibfnamefont {M.}~\bibnamefont
  {Hase}}, \bibinfo {author} {\bibfnamefont {H.}~\bibnamefont {Kuroe}},
  \bibinfo {author} {\bibfnamefont {K.}~\bibnamefont {Ozawa}}, \bibinfo
  {author} {\bibfnamefont {O.}~\bibnamefont {Suzuki}}, \bibinfo {author}
  {\bibfnamefont {H.}~\bibnamefont {Kitazawa}}, \bibinfo {author}
  {\bibfnamefont {G.}~\bibnamefont {Kido}}, \ and\ \bibinfo {author}
  {\bibfnamefont {T.}~\bibnamefont {Sekine}},\ }\href {\doibase
  10.1103/PhysRevB.70.104426} {\bibfield  {journal} {\bibinfo  {journal} {Phys.
  Rev. B}\ }\textbf {\bibinfo {volume} {70}},\ \bibinfo {pages} {104426}
  (\bibinfo {year} {2004})}\BibitemShut {NoStop}%
\bibitem [{\citenamefont {Hase}\ \emph {et~al.}(2005)\citenamefont {Hase},
  \citenamefont {Ozawa}, \citenamefont {Suzuki}, \citenamefont {Kitazawa},
  \citenamefont {Kido}, \citenamefont {Kuroe},\ and\ \citenamefont
  {Sekine}}]{Hase2005}%
  \BibitemOpen
  \bibfield  {author} {\bibinfo {author} {\bibfnamefont {M.}~\bibnamefont
  {Hase}}, \bibinfo {author} {\bibfnamefont {K.}~\bibnamefont {Ozawa}},
  \bibinfo {author} {\bibfnamefont {O.}~\bibnamefont {Suzuki}}, \bibinfo
  {author} {\bibfnamefont {H.}~\bibnamefont {Kitazawa}}, \bibinfo {author}
  {\bibfnamefont {G.}~\bibnamefont {Kido}}, \bibinfo {author} {\bibfnamefont
  {H.}~\bibnamefont {Kuroe}}, \ and\ \bibinfo {author} {\bibfnamefont
  {T.}~\bibnamefont {Sekine}},\ }\href {\doibase 10.1063/1.1851675} {\bibfield
  {journal} {\bibinfo  {journal} {J. Appl. Phys.}\ }\textbf {\bibinfo {volume}
  {97}},\ \bibinfo {pages} {10B303} (\bibinfo {year} {2005})}\BibitemShut
  {NoStop}%
\bibitem [{\citenamefont {Yasui}\ \emph {et~al.}(2014)\citenamefont {Yasui},
  \citenamefont {Okazaki}, \citenamefont {Terasaki}, \citenamefont {Hase},
  \citenamefont {Hagihala}, \citenamefont {Masuda},\ and\ \citenamefont
  {Sakakibara}}]{Yasui2014}%
  \BibitemOpen
  \bibfield  {author} {\bibinfo {author} {\bibfnamefont {Y.}~\bibnamefont
  {Yasui}}, \bibinfo {author} {\bibfnamefont {R.}~\bibnamefont {Okazaki}},
  \bibinfo {author} {\bibfnamefont {I.}~\bibnamefont {Terasaki}}, \bibinfo
  {author} {\bibfnamefont {M.}~\bibnamefont {Hase}}, \bibinfo {author}
  {\bibfnamefont {M.}~\bibnamefont {Hagihala}}, \bibinfo {author}
  {\bibfnamefont {T.}~\bibnamefont {Masuda}}, \ and\ \bibinfo {author}
  {\bibfnamefont {T.}~\bibnamefont {Sakakibara}},\ }\href {\doibase
  10.7566/JPSCP.3.014014} {\bibfield  {journal} {\bibinfo  {journal} {JPS Conf.
  Proc.}\ }\textbf {\bibinfo {volume} {3}},\ \bibinfo {pages} {014014}
  (\bibinfo {year} {2014})}\BibitemShut {NoStop}%
\bibitem [{\citenamefont {Yasui}\ \emph
  {et~al.}(2013{\natexlab{a}})\citenamefont {Yasui}, \citenamefont
  {Yanagisawa}, \citenamefont {Okazaki},\ and\ \citenamefont
  {Terasaki}}]{Yasui2013_1}%
  \BibitemOpen
  \bibfield  {author} {\bibinfo {author} {\bibfnamefont {Y.}~\bibnamefont
  {Yasui}}, \bibinfo {author} {\bibfnamefont {Y.}~\bibnamefont {Yanagisawa}},
  \bibinfo {author} {\bibfnamefont {R.}~\bibnamefont {Okazaki}}, \ and\
  \bibinfo {author} {\bibfnamefont {I.}~\bibnamefont {Terasaki}},\ }\href
  {\doibase 10.1103/PhysRevB.87.054411} {\bibfield  {journal} {\bibinfo
  {journal} {Phys. Rev. B}\ }\textbf {\bibinfo {volume} {87}},\ \bibinfo
  {pages} {054411} (\bibinfo {year} {2013}{\natexlab{a}})}\BibitemShut
  {NoStop}%
\bibitem [{\citenamefont {Yasui}\ \emph
  {et~al.}(2013{\natexlab{b}})\citenamefont {Yasui}, \citenamefont
  {Yanagisawa}, \citenamefont {Okazaki}, \citenamefont {Terasaki},
  \citenamefont {Yamaguchi},\ and\ \citenamefont {Kimura}}]{Yasui2013_2}%
  \BibitemOpen
  \bibfield  {author} {\bibinfo {author} {\bibfnamefont {Y.}~\bibnamefont
  {Yasui}}, \bibinfo {author} {\bibfnamefont {Y.}~\bibnamefont {Yanagisawa}},
  \bibinfo {author} {\bibfnamefont {R.}~\bibnamefont {Okazaki}}, \bibinfo
  {author} {\bibfnamefont {I.}~\bibnamefont {Terasaki}}, \bibinfo {author}
  {\bibfnamefont {Y.}~\bibnamefont {Yamaguchi}}, \ and\ \bibinfo {author}
  {\bibfnamefont {T.}~\bibnamefont {Kimura}},\ }\href@noop {} {\bibfield
  {journal} {\bibinfo  {journal} {Journal of Applied Physics}\ }\textbf
  {\bibinfo {volume} {113}},\ \bibinfo {pages} {17D910} (\bibinfo {year}
  {2013}{\natexlab{b}})}\BibitemShut {NoStop}%
\bibitem [{\citenamefont {Reynolds}\ \emph {et~al.}(2019)\citenamefont
  {Reynolds}, \citenamefont {Mannig}, \citenamefont {Luetkens}, \citenamefont
  {Baines}, \citenamefont {Goko}, \citenamefont {Scheuermann}, \citenamefont
  {Keller}, \citenamefont {Bartkowiak}, \citenamefont {Fujimura}, \citenamefont
  {Yasui}, \citenamefont {Niedermayer},\ and\ \citenamefont
  {White}}]{Reynolds2019}%
  \BibitemOpen
  \bibfield  {author} {\bibinfo {author} {\bibfnamefont {N.}~\bibnamefont
  {Reynolds}}, \bibinfo {author} {\bibfnamefont {A.}~\bibnamefont {Mannig}},
  \bibinfo {author} {\bibfnamefont {H.}~\bibnamefont {Luetkens}}, \bibinfo
  {author} {\bibfnamefont {C.}~\bibnamefont {Baines}}, \bibinfo {author}
  {\bibfnamefont {T.}~\bibnamefont {Goko}}, \bibinfo {author} {\bibfnamefont
  {R.}~\bibnamefont {Scheuermann}}, \bibinfo {author} {\bibfnamefont
  {L.}~\bibnamefont {Keller}}, \bibinfo {author} {\bibfnamefont
  {M.}~\bibnamefont {Bartkowiak}}, \bibinfo {author} {\bibfnamefont
  {A.}~\bibnamefont {Fujimura}}, \bibinfo {author} {\bibfnamefont
  {Y.}~\bibnamefont {Yasui}}, \bibinfo {author} {\bibfnamefont
  {C.}~\bibnamefont {Niedermayer}}, \ and\ \bibinfo {author} {\bibfnamefont
  {J.~S.}\ \bibnamefont {White}},\ }\href {\doibase 10.1103/PhysRevB.99.214443}
  {\bibfield  {journal} {\bibinfo  {journal} {Phys. Rev. B}\ }\textbf {\bibinfo
  {volume} {99}},\ \bibinfo {pages} {214443} (\bibinfo {year}
  {2019})}\BibitemShut {NoStop}%
\bibitem [{\citenamefont {Kuroe}\ \emph {et~al.}(2006)\citenamefont {Kuroe},
  \citenamefont {Hamasaki}, \citenamefont {Sekine}, \citenamefont {Hase},
  \citenamefont {Naka},\ and\ \citenamefont {Maeshima}}]{Kuroe2006}%
  \BibitemOpen
  \bibfield  {author} {\bibinfo {author} {\bibfnamefont {H.}~\bibnamefont
  {Kuroe}}, \bibinfo {author} {\bibfnamefont {T.}~\bibnamefont {Hamasaki}},
  \bibinfo {author} {\bibfnamefont {T.}~\bibnamefont {Sekine}}, \bibinfo
  {author} {\bibfnamefont {M.}~\bibnamefont {Hase}}, \bibinfo {author}
  {\bibfnamefont {T.}~\bibnamefont {Naka}}, \ and\ \bibinfo {author}
  {\bibfnamefont {N.}~\bibnamefont {Maeshima}},\ }\href@noop {} {\bibfield
  {journal} {\bibinfo  {journal} {AIP Conf. Proc.}\ }\textbf {\bibinfo {volume}
  {850}},\ \bibinfo {pages} {1049} (\bibinfo {year} {2006})}\BibitemShut
  {NoStop}%
\bibitem [{\citenamefont {Hamasaki}\ \emph {et~al.}(2007)\citenamefont
  {Hamasaki}, \citenamefont {Kuroe}, \citenamefont {Sekine}, \citenamefont
  {Naka}, \citenamefont {Hase}, \citenamefont {Maeshima}, \citenamefont
  {Saiga},\ and\ \citenamefont {Uwatoko}}]{Hamasaki2007}%
  \BibitemOpen
  \bibfield  {author} {\bibinfo {author} {\bibfnamefont {T.}~\bibnamefont
  {Hamasaki}}, \bibinfo {author} {\bibfnamefont {H.}~\bibnamefont {Kuroe}},
  \bibinfo {author} {\bibfnamefont {T.}~\bibnamefont {Sekine}}, \bibinfo
  {author} {\bibfnamefont {T.}~\bibnamefont {Naka}}, \bibinfo {author}
  {\bibfnamefont {M.}~\bibnamefont {Hase}}, \bibinfo {author} {\bibfnamefont
  {N.}~\bibnamefont {Maeshima}}, \bibinfo {author} {\bibfnamefont
  {Y.}~\bibnamefont {Saiga}}, \ and\ \bibinfo {author} {\bibfnamefont
  {Y.}~\bibnamefont {Uwatoko}},\ }\href {\doibase
  https://doi.org/10.1016/j.jmmm.2006.10.370} {\bibfield  {journal} {\bibinfo
  {journal} {Journal of Magnetism and Magnetic Materials}\ }\textbf {\bibinfo
  {volume} {310}},\ \bibinfo {pages} {e394 } (\bibinfo {year} {2007})},\
  \bibinfo {note} {proceedings of the 17th International Conference on
  Magnetism}\BibitemShut {NoStop}%
\bibitem [{\citenamefont {Yagi}\ \emph {et~al.}(2017)\citenamefont {Yagi},
  \citenamefont {Matsui}, \citenamefont {Goto}, \citenamefont {Hase},\ and\
  \citenamefont {Sasaki}}]{Yagi2017}%
  \BibitemOpen
  \bibfield  {author} {\bibinfo {author} {\bibfnamefont {A.}~\bibnamefont
  {Yagi}}, \bibinfo {author} {\bibfnamefont {K.}~\bibnamefont {Matsui}},
  \bibinfo {author} {\bibfnamefont {T.}~\bibnamefont {Goto}}, \bibinfo {author}
  {\bibfnamefont {M.}~\bibnamefont {Hase}}, \ and\ \bibinfo {author}
  {\bibfnamefont {T.}~\bibnamefont {Sasaki}},\ }\href@noop {} {\bibfield
  {journal} {\bibinfo  {journal} {J. Phys.: Conf. Ser.}\ }\textbf {\bibinfo
  {volume} {828}},\ \bibinfo {pages} {012016} (\bibinfo {year}
  {2017})}\BibitemShut {NoStop}%
\bibitem [{\citenamefont {Matsui}\ \emph {et~al.}(2017)\citenamefont {Matsui},
  \citenamefont {Yagi}, \citenamefont {Hoshino}, \citenamefont {Atarashi},
  \citenamefont {Hase}, \citenamefont {Sasaki},\ and\ \citenamefont
  {Goto}}]{Matsui2017}%
  \BibitemOpen
  \bibfield  {author} {\bibinfo {author} {\bibfnamefont {K.}~\bibnamefont
  {Matsui}}, \bibinfo {author} {\bibfnamefont {A.}~\bibnamefont {Yagi}},
  \bibinfo {author} {\bibfnamefont {Y.}~\bibnamefont {Hoshino}}, \bibinfo
  {author} {\bibfnamefont {S.}~\bibnamefont {Atarashi}}, \bibinfo {author}
  {\bibfnamefont {M.}~\bibnamefont {Hase}}, \bibinfo {author} {\bibfnamefont
  {T.}~\bibnamefont {Sasaki}}, \ and\ \bibinfo {author} {\bibfnamefont
  {T.}~\bibnamefont {Goto}},\ }\href {\doibase 10.1103/PhysRevB.96.220402}
  {\bibfield  {journal} {\bibinfo  {journal} {Phys. Rev. B}\ }\textbf {\bibinfo
  {volume} {96}},\ \bibinfo {pages} {220402(R)} (\bibinfo {year}
  {2017})}\BibitemShut {NoStop}%
\bibitem [{\citenamefont {Tomiyasu}\ \emph {et~al.}(2009)\citenamefont
  {Tomiyasu}, \citenamefont {Fujita}, \citenamefont {Kolesnikov}, \citenamefont
  {Bewley}, \citenamefont {Bull},\ and\ \citenamefont
  {Bennington}}]{Tomiyasu2009}%
  \BibitemOpen
  \bibfield  {author} {\bibinfo {author} {\bibfnamefont {K.}~\bibnamefont
  {Tomiyasu}}, \bibinfo {author} {\bibfnamefont {M.}~\bibnamefont {Fujita}},
  \bibinfo {author} {\bibfnamefont {A.~I.}\ \bibnamefont {Kolesnikov}},
  \bibinfo {author} {\bibfnamefont {R.~I.}\ \bibnamefont {Bewley}}, \bibinfo
  {author} {\bibfnamefont {M.}~\bibnamefont {Bull}}, \ and\ \bibinfo {author}
  {\bibfnamefont {S.}~\bibnamefont {Bennington}},\ }\href@noop {} {\bibfield
  {journal} {\bibinfo  {journal} {Applied Physics Letters}\ }\textbf {\bibinfo
  {volume} {94}},\ \bibinfo {pages} {092502} (\bibinfo {year}
  {2009})}\BibitemShut {NoStop}%
\bibitem [{\citenamefont {Ohira-Kawamura}\ \emph {et~al.}(2018)\citenamefont
  {Ohira-Kawamura}, \citenamefont {Tomiyasu}, \citenamefont {Koda},
  \citenamefont {Sari}, \citenamefont {Asih}, \citenamefont {Yoon},
  \citenamefont {Watanabe},\ and\ \citenamefont {Nakajima}}]{Kawamura2018}%
  \BibitemOpen
  \bibfield  {author} {\bibinfo {author} {\bibfnamefont {S.}~\bibnamefont
  {Ohira-Kawamura}}, \bibinfo {author} {\bibfnamefont {K.}~\bibnamefont
  {Tomiyasu}}, \bibinfo {author} {\bibfnamefont {A.}~\bibnamefont {Koda}},
  \bibinfo {author} {\bibfnamefont {D.~P.}\ \bibnamefont {Sari}}, \bibinfo
  {author} {\bibfnamefont {R.}~\bibnamefont {Asih}}, \bibinfo {author}
  {\bibfnamefont {S.}~\bibnamefont {Yoon}}, \bibinfo {author} {\bibfnamefont
  {I.}~\bibnamefont {Watanabe}}, \ and\ \bibinfo {author} {\bibfnamefont
  {K.}~\bibnamefont {Nakajima}},\ }\href {\doibase 10.7566/JPSCP.21.011007}
  {\bibfield  {journal} {\bibinfo  {journal} {JPS Conf. Proc.}\ }\textbf
  {\bibinfo {volume} {21}},\ \bibinfo {pages} {011007} (\bibinfo {year}
  {2018})}\BibitemShut {NoStop}%
\bibitem [{\citenamefont {Katsura}\ \emph {et~al.}(2005)\citenamefont
  {Katsura}, \citenamefont {Nagaosa},\ and\ \citenamefont
  {Balatsky}}]{Katsura2005}%
  \BibitemOpen
  \bibfield  {author} {\bibinfo {author} {\bibfnamefont {H.}~\bibnamefont
  {Katsura}}, \bibinfo {author} {\bibfnamefont {N.}~\bibnamefont {Nagaosa}}, \
  and\ \bibinfo {author} {\bibfnamefont {A.~V.}\ \bibnamefont {Balatsky}},\
  }\href {\doibase 10.1103/PhysRevLett.95.057205} {\bibfield  {journal}
  {\bibinfo  {journal} {Phys. Rev. Lett.}\ }\textbf {\bibinfo {volume} {95}},\
  \bibinfo {pages} {057205} (\bibinfo {year} {2005})}\BibitemShut {NoStop}%
\bibitem [{\citenamefont {Mostovoy}(2006)}]{Mostovoy2006}%
  \BibitemOpen
  \bibfield  {author} {\bibinfo {author} {\bibfnamefont {M.}~\bibnamefont
  {Mostovoy}},\ }\href {\doibase 10.1103/PhysRevLett.96.067601} {\bibfield
  {journal} {\bibinfo  {journal} {Phys. Rev. Lett.}\ }\textbf {\bibinfo
  {volume} {96}},\ \bibinfo {pages} {067601} (\bibinfo {year}
  {2006})}\BibitemShut {NoStop}%
\bibitem [{\citenamefont {Jia}\ \emph {et~al.}(2007)\citenamefont {Jia},
  \citenamefont {Onoda}, \citenamefont {Nagaosa},\ and\ \citenamefont
  {Han}}]{Jia2007}%
  \BibitemOpen
  \bibfield  {author} {\bibinfo {author} {\bibfnamefont {C.}~\bibnamefont
  {Jia}}, \bibinfo {author} {\bibfnamefont {S.}~\bibnamefont {Onoda}}, \bibinfo
  {author} {\bibfnamefont {N.}~\bibnamefont {Nagaosa}}, \ and\ \bibinfo
  {author} {\bibfnamefont {J.~H.}\ \bibnamefont {Han}},\ }\href {\doibase
  10.1103/PhysRevB.76.144424} {\bibfield  {journal} {\bibinfo  {journal} {Phys.
  Rev. B}\ }\textbf {\bibinfo {volume} {76}},\ \bibinfo {pages} {144424}
  (\bibinfo {year} {2007})}\BibitemShut {NoStop}%
\bibitem [{\citenamefont {Xiang}\ and\ \citenamefont
  {Whangbo}(2007)}]{Xiang2007}%
  \BibitemOpen
  \bibfield  {author} {\bibinfo {author} {\bibfnamefont {H.~J.}\ \bibnamefont
  {Xiang}}\ and\ \bibinfo {author} {\bibfnamefont {M.-H.}\ \bibnamefont
  {Whangbo}},\ }\href {\doibase 10.1103/PhysRevLett.99.257203} {\bibfield
  {journal} {\bibinfo  {journal} {Phys. Rev. Lett.}\ }\textbf {\bibinfo
  {volume} {99}},\ \bibinfo {pages} {257203} (\bibinfo {year}
  {2007})}\BibitemShut {NoStop}%
\bibitem [{\citenamefont {Zvyagin}\ \emph {et~al.}(2004)\citenamefont
  {Zvyagin}, \citenamefont {Krzystek}, \citenamefont {van Loosdrecht},
  \citenamefont {Dhalenne},\ and\ \citenamefont {Revcolevschi}}]{Zvyagin2004}%
  \BibitemOpen
  \bibfield  {author} {\bibinfo {author} {\bibfnamefont {S.}~\bibnamefont
  {Zvyagin}}, \bibinfo {author} {\bibfnamefont {J.}~\bibnamefont {Krzystek}},
  \bibinfo {author} {\bibfnamefont {P.}~\bibnamefont {van Loosdrecht}},
  \bibinfo {author} {\bibfnamefont {G.}~\bibnamefont {Dhalenne}}, \ and\
  \bibinfo {author} {\bibfnamefont {A.}~\bibnamefont {Revcolevschi}},\ }\href
  {\doibase https://doi.org/10.1016/j.physb.2004.01.009} {\bibfield  {journal}
  {\bibinfo  {journal} {Physica B: Condensed Matter}\ }\textbf {\bibinfo
  {volume} {346-347}},\ \bibinfo {pages} {1 } (\bibinfo {year}
  {2004})}\BibitemShut {NoStop}%
\bibitem [{\citenamefont {Zapf}\ \emph {et~al.}(2014)\citenamefont {Zapf},
  \citenamefont {Jaime},\ and\ \citenamefont {Batista}}]{Zapf2014}%
  \BibitemOpen
  \bibfield  {author} {\bibinfo {author} {\bibfnamefont {V.}~\bibnamefont
  {Zapf}}, \bibinfo {author} {\bibfnamefont {M.}~\bibnamefont {Jaime}}, \ and\
  \bibinfo {author} {\bibfnamefont {C.~D.}\ \bibnamefont {Batista}},\ }\href
  {\doibase 10.1103/RevModPhys.86.563} {\bibfield  {journal} {\bibinfo
  {journal} {Rev. Mod. Phys.}\ }\textbf {\bibinfo {volume} {86}},\ \bibinfo
  {pages} {563} (\bibinfo {year} {2014})}\BibitemShut {NoStop}%
\bibitem [{\citenamefont {Korepin}\ and\ \citenamefont
  {Slavnov}(1990)}]{Korepin1990}%
  \BibitemOpen
  \bibfield  {author} {\bibinfo {author} {\bibfnamefont {V.~E.}\ \bibnamefont
  {Korepin}}\ and\ \bibinfo {author} {\bibfnamefont {N.~A.}\ \bibnamefont
  {Slavnov}},\ }\href {\doibase 10.1007/BF02096781} {\bibfield  {journal}
  {\bibinfo  {journal} {Commun. Math. Phys.}\ }\textbf {\bibinfo {volume}
  {129}},\ \bibinfo {pages} {103} (\bibinfo {year} {1990})}\BibitemShut
  {NoStop}%
\bibitem [{\citenamefont {Sachdev}\ \emph {et~al.}(1994)\citenamefont
  {Sachdev}, \citenamefont {Senthil},\ and\ \citenamefont
  {Shankar}}]{Sachdev1994}%
  \BibitemOpen
  \bibfield  {author} {\bibinfo {author} {\bibfnamefont {S.}~\bibnamefont
  {Sachdev}}, \bibinfo {author} {\bibfnamefont {T.}~\bibnamefont {Senthil}}, \
  and\ \bibinfo {author} {\bibfnamefont {R.}~\bibnamefont {Shankar}},\ }\href
  {\doibase 10.1103/PhysRevB.50.258} {\bibfield  {journal} {\bibinfo  {journal}
  {Phys. Rev. B}\ }\textbf {\bibinfo {volume} {50}},\ \bibinfo {pages} {258}
  (\bibinfo {year} {1994})}\BibitemShut {NoStop}%
\bibitem [{\citenamefont {R\"uegg}\ \emph {et~al.}(2008)\citenamefont
  {R\"uegg}, \citenamefont {Kiefer}, \citenamefont {Thielemann}, \citenamefont
  {McMorrow}, \citenamefont {Zapf}, \citenamefont {Normand}, \citenamefont
  {Zvonarev}, \citenamefont {Bouillot}, \citenamefont {Kollath}, \citenamefont
  {Giamarchi}, \citenamefont {Capponi}, \citenamefont {Poilblanc},
  \citenamefont {Biner},\ and\ \citenamefont {Kr\"amer}}]{Ruegg2008}%
  \BibitemOpen
  \bibfield  {author} {\bibinfo {author} {\bibfnamefont {C.}~\bibnamefont
  {R\"uegg}}, \bibinfo {author} {\bibfnamefont {K.}~\bibnamefont {Kiefer}},
  \bibinfo {author} {\bibfnamefont {B.}~\bibnamefont {Thielemann}}, \bibinfo
  {author} {\bibfnamefont {D.~F.}\ \bibnamefont {McMorrow}}, \bibinfo {author}
  {\bibfnamefont {V.}~\bibnamefont {Zapf}}, \bibinfo {author} {\bibfnamefont
  {B.}~\bibnamefont {Normand}}, \bibinfo {author} {\bibfnamefont {M.~B.}\
  \bibnamefont {Zvonarev}}, \bibinfo {author} {\bibfnamefont {P.}~\bibnamefont
  {Bouillot}}, \bibinfo {author} {\bibfnamefont {C.}~\bibnamefont {Kollath}},
  \bibinfo {author} {\bibfnamefont {T.}~\bibnamefont {Giamarchi}}, \bibinfo
  {author} {\bibfnamefont {S.}~\bibnamefont {Capponi}}, \bibinfo {author}
  {\bibfnamefont {D.}~\bibnamefont {Poilblanc}}, \bibinfo {author}
  {\bibfnamefont {D.}~\bibnamefont {Biner}}, \ and\ \bibinfo {author}
  {\bibfnamefont {K.~W.}\ \bibnamefont {Kr\"amer}},\ }\href {\doibase
  10.1103/PhysRevLett.101.247202} {\bibfield  {journal} {\bibinfo  {journal}
  {Phys. Rev. Lett.}\ }\textbf {\bibinfo {volume} {101}},\ \bibinfo {pages}
  {247202} (\bibinfo {year} {2008})}\BibitemShut {NoStop}%
\bibitem [{\citenamefont {Breunig}\ \emph {et~al.}(2017)\citenamefont
  {Breunig}, \citenamefont {Garst}, \citenamefont {Kl{\"u}mper}, \citenamefont
  {Rohrkamp}, \citenamefont {Turnbull},\ and\ \citenamefont
  {Lorenz}}]{Breunig2017}%
  \BibitemOpen
  \bibfield  {author} {\bibinfo {author} {\bibfnamefont {O.}~\bibnamefont
  {Breunig}}, \bibinfo {author} {\bibfnamefont {M.}~\bibnamefont {Garst}},
  \bibinfo {author} {\bibfnamefont {A.}~\bibnamefont {Kl{\"u}mper}}, \bibinfo
  {author} {\bibfnamefont {J.}~\bibnamefont {Rohrkamp}}, \bibinfo {author}
  {\bibfnamefont {M.~M.}\ \bibnamefont {Turnbull}}, \ and\ \bibinfo {author}
  {\bibfnamefont {T.}~\bibnamefont {Lorenz}},\ }\href
  {https://advances.sciencemag.org/content/3/12/eaao3773} {\bibfield  {journal}
  {\bibinfo  {journal} {Science Advances}\ }\textbf {\bibinfo {volume} {3}},\ \bibinfo {pages} {eaao3773}
  (\bibinfo {year} {2017})}\BibitemShut {NoStop}%
\bibitem [{\citenamefont {Blosser}\ \emph {et~al.}(2018)\citenamefont
  {Blosser}, \citenamefont {Bhartiya}, \citenamefont {Voneshen},\ and\
  \citenamefont {Zheludev}}]{Blosser2018}%
  \BibitemOpen
  \bibfield  {author} {\bibinfo {author} {\bibfnamefont {D.}~\bibnamefont
  {Blosser}}, \bibinfo {author} {\bibfnamefont {V.~K.}\ \bibnamefont
  {Bhartiya}}, \bibinfo {author} {\bibfnamefont {D.~J.}\ \bibnamefont
  {Voneshen}}, \ and\ \bibinfo {author} {\bibfnamefont {A.}~\bibnamefont
  {Zheludev}},\ }\href {\doibase 10.1103/PhysRevLett.121.247201} {\bibfield
  {journal} {\bibinfo  {journal} {Phys. Rev. Lett.}\ }\textbf {\bibinfo
  {volume} {121}},\ \bibinfo {pages} {247201} (\bibinfo {year}
  {2018})}\BibitemShut {NoStop}%
\bibitem [{\citenamefont {Troyer}\ \emph {et~al.}(1994)\citenamefont {Troyer},
  \citenamefont {Tsunetsugu},\ and\ \citenamefont {W\"urtz}}]{Troyer1994}%
  \BibitemOpen
  \bibfield  {author} {\bibinfo {author} {\bibfnamefont {M.}~\bibnamefont
  {Troyer}}, \bibinfo {author} {\bibfnamefont {H.}~\bibnamefont {Tsunetsugu}},
  \ and\ \bibinfo {author} {\bibfnamefont {D.}~\bibnamefont {W\"urtz}},\ }\href
  {\doibase 10.1103/PhysRevB.50.13515} {\bibfield  {journal} {\bibinfo
  {journal} {Phys. Rev. B}\ }\textbf {\bibinfo {volume} {50}},\ \bibinfo
  {pages} {13515} (\bibinfo {year} {1994})}\BibitemShut {NoStop}%
\bibitem [{\citenamefont {Hong}\ \emph {et~al.}(2010)\citenamefont {Hong},
  \citenamefont {Kim}, \citenamefont {Hotta}, \citenamefont {Takano},
  \citenamefont {Tremelling}, \citenamefont {Turnbull}, \citenamefont {Landee},
  \citenamefont {Kang}, \citenamefont {Christensen}, \citenamefont {Lefmann},
  \citenamefont {Schmidt}, \citenamefont {Uhrig},\ and\ \citenamefont
  {Broholm}}]{Hong2010}%
  \BibitemOpen
  \bibfield  {author} {\bibinfo {author} {\bibfnamefont {T.}~\bibnamefont
  {Hong}}, \bibinfo {author} {\bibfnamefont {Y.~H.}\ \bibnamefont {Kim}},
  \bibinfo {author} {\bibfnamefont {C.}~\bibnamefont {Hotta}}, \bibinfo
  {author} {\bibfnamefont {Y.}~\bibnamefont {Takano}}, \bibinfo {author}
  {\bibfnamefont {G.}~\bibnamefont {Tremelling}}, \bibinfo {author}
  {\bibfnamefont {M.~M.}\ \bibnamefont {Turnbull}}, \bibinfo {author}
  {\bibfnamefont {C.~P.}\ \bibnamefont {Landee}}, \bibinfo {author}
  {\bibfnamefont {H.-J.}\ \bibnamefont {Kang}}, \bibinfo {author}
  {\bibfnamefont {N.~B.}\ \bibnamefont {Christensen}}, \bibinfo {author}
  {\bibfnamefont {K.}~\bibnamefont {Lefmann}}, \bibinfo {author} {\bibfnamefont
  {K.~P.}\ \bibnamefont {Schmidt}}, \bibinfo {author} {\bibfnamefont {G.~S.}\
  \bibnamefont {Uhrig}}, \ and\ \bibinfo {author} {\bibfnamefont
  {C.}~\bibnamefont {Broholm}},\ }\href {\doibase
  10.1103/PhysRevLett.105.137207} {\bibfield  {journal} {\bibinfo  {journal}
  {Phys. Rev. Lett.}\ }\textbf {\bibinfo {volume} {105}},\ \bibinfo {pages}
  {137207} (\bibinfo {year} {2010})}\BibitemShut {NoStop}%
\bibitem [{\citenamefont {Continentino}(2017)}]{Continentino2017}%
  \BibitemOpen
  \bibfield  {author} {\bibinfo {author} {\bibfnamefont {M.}~\bibnamefont
  {Continentino}},\ }\href@noop {} {\emph {\bibinfo {title} {Quantum scaling in
  many-body systems}}}\ (\bibinfo  {publisher} {Cambridge University Press, Cambridge},\
  \bibinfo {year} {2017})\BibitemShut {NoStop}%
\bibitem [{\citenamefont {Bonner}\ and\ \citenamefont
  {Fisher}(1964)}]{Bonner1964}%
  \BibitemOpen
  \bibfield  {author} {\bibinfo {author} {\bibfnamefont {J.~C.}\ \bibnamefont
  {Bonner}}\ and\ \bibinfo {author} {\bibfnamefont {M.~E.}\ \bibnamefont
  {Fisher}},\ }\href {\doibase 10.1103/PhysRev.135.A640} {\bibfield  {journal}
  {\bibinfo  {journal} {Phys. Rev.}\ }\textbf {\bibinfo {volume} {135}},\
  \bibinfo {pages} {A640} (\bibinfo {year} {1964})}\BibitemShut {NoStop}%
\bibitem [{\citenamefont {Batyev}\ and\ \citenamefont
  {Braginski}(1984)}]{Batyev1984}%
  \BibitemOpen
  \bibfield  {author} {\bibinfo {author} {\bibfnamefont {E.~G.}\ \bibnamefont
  {Batyev}}\ and\ \bibinfo {author} {\bibfnamefont {L.~S.}\ \bibnamefont
  {Braginski}},\ }\href@noop {} {\bibfield  {journal} {\bibinfo  {journal}
  {Sov. Phys. JETP}\ }\textbf {\bibinfo {volume} {60}},\ \bibinfo {pages} {781}
  (\bibinfo {year} {1984})}\BibitemShut {NoStop}%
\bibitem [{\citenamefont {Giamarchi}\ and\ \citenamefont
  {Tsvelik}(1999)}]{Giamarchi1999}%
  \BibitemOpen
  \bibfield  {author} {\bibinfo {author} {\bibfnamefont {T.}~\bibnamefont
  {Giamarchi}}\ and\ \bibinfo {author} {\bibfnamefont {A.~M.}\ \bibnamefont
  {Tsvelik}},\ }\href {\doibase 10.1103/PhysRevB.59.11398} {\bibfield
  {journal} {\bibinfo  {journal} {Phys. Rev. B}\ }\textbf {\bibinfo {volume}
  {59}},\ \bibinfo {pages} {11398} (\bibinfo {year} {1999})}\BibitemShut
  {NoStop}%
\bibitem [{\citenamefont {Svistov}\ \emph {et~al.}(2010)\citenamefont
  {Svistov}, \citenamefont {Prozorova}, \citenamefont {Bush},\ and\
  \citenamefont {Kamentsev}}]{Svistov2010}%
  \BibitemOpen
  \bibfield  {author} {\bibinfo {author} {\bibfnamefont {L.~E.}\ \bibnamefont
  {Svistov}}, \bibinfo {author} {\bibfnamefont {L.~A.}\ \bibnamefont
  {Prozorova}}, \bibinfo {author} {\bibfnamefont {A.~A.}\ \bibnamefont {Bush}},
  \ and\ \bibinfo {author} {\bibfnamefont {K.~E.}\ \bibnamefont {Kamentsev}},\
  }\href {\doibase 10.1088/1742-6596/200/2/022062} {\bibfield  {journal}
  {\bibinfo  {journal} {J. Phys.: Conf. Ser.}\ }\textbf {\bibinfo {volume}
  {200}},\ \bibinfo {pages} {022062} (\bibinfo {year} {2010})}\BibitemShut
  {NoStop}%
\bibitem [{\citenamefont {Fujita}\ \emph {et~al.}(2014)\citenamefont {Fujita},
  \citenamefont {Hagiwara}, \citenamefont {Inada}, \citenamefont {Yasui},\ and\
  \citenamefont {Terasaki}}]{Fujita2014}%
  \BibitemOpen
  \bibfield  {author} {\bibinfo {author} {\bibfnamefont {T.}~\bibnamefont
  {Fujita}}, \bibinfo {author} {\bibfnamefont {M.}~\bibnamefont {Hagiwara}},
  \bibinfo {author} {\bibfnamefont {M.}~\bibnamefont {Inada}}, \bibinfo
  {author} {\bibfnamefont {Y.}~\bibnamefont {Yasui}}, \ and\ \bibinfo {author}
  {\bibfnamefont {I.}~\bibnamefont {Terasaki}},\ }\href {\doibase
  10.7566/JPSCP.3.014028} {\bibfield  {journal} {\bibinfo  {journal} {JPS Conf.
  Proc.}\ }\textbf {\bibinfo {volume} {3}},\ \bibinfo {pages} {014028}
  (\bibinfo {year} {2014})}\BibitemShut {NoStop}%
\bibitem [{\citenamefont {Prozorova}\ \emph {et~al.}(2016)\citenamefont
  {Prozorova}, \citenamefont {Svistov}, \citenamefont {Vasiliev},\ and\
  \citenamefont {Prokofiev}}]{Prozorova2016}%
  \BibitemOpen
  \bibfield  {author} {\bibinfo {author} {\bibfnamefont {L.~A.}\ \bibnamefont
  {Prozorova}}, \bibinfo {author} {\bibfnamefont {L.~E.}\ \bibnamefont
  {Svistov}}, \bibinfo {author} {\bibfnamefont {A.~M.}\ \bibnamefont
  {Vasiliev}}, \ and\ \bibinfo {author} {\bibfnamefont {A.}~\bibnamefont
  {Prokofiev}},\ }\href {\doibase 10.1103/PhysRevB.94.224402} {\bibfield
  {journal} {\bibinfo  {journal} {Phys. Rev. B}\ }\textbf {\bibinfo {volume}
  {94}},\ \bibinfo {pages} {224402} (\bibinfo {year} {2016})}\BibitemShut
  {NoStop}%
\end{thebibliography}
%

\end{document}